\documentclass[review]{elsarticle}

\usepackage{lineno,hyperref,graphicx,amsmath,amssymb,bm,lscape}
\modulolinenumbers[5]

\journal{Journal of \LaTeX\ Templates}

\begin{document}

\begin{frontmatter}

\title{Analysing supersymmetric transformed $\alpha$-nucleus potentials
  with electric-multipole transitions}

\author{T. Arai, W. Horiuchi*}
\address{Department of Physics, Hokkaido University, Sapporo 060-0810, Japan}
\cortext[mycorrespondingauthor]{Corresponding author}

\author{D. Baye}
\address{Physique Quantique, and Physique Nucl\'eaire Th\'eorique et Physique Math\'ematique, C.P. 229,
  Universit\'e Libre de Bruxelles (ULB), B-1050 Brussels, Belgium}



\begin{abstract}
Alpha($^{4}$He)-cluster models have often been used
to describe light nuclei. Towards the application
to multi-cluster systems involving heavy clusters,
we study the relative wave functions of the
$\alpha+^{16}$O and $\alpha+^{40}$Ca systems
generated from phase-shift-equivalent potentials. 
In general, a potential between clusters is deep
accommodating several redundant bound states
which should be removed in an appropriate way.
To avoid such a complicated computation,
we generate a shallow-singular potential
by using supersymmetric transformations
from the original deep potential.
Changes in the relative wave functions by the transformations
are quantified with electric-multipole transitions which give
a different radial sensitivity to the wave function
depending on their multipolarity.
Despite the fact that the original and transformed potentials
give exactly the same phase shift,
some observables are unfavorably modified. 
A possible way to obtain a desired supersymmetric potential
is proposed.  
\end{abstract}


\end{frontmatter}


\section{Introduction}

Nuclear clustering is one of the characteristic phenomena
in atomic nuclear systems. Since a $^{4}$He nucleus ($\alpha$)
is tightly bound by about 28 MeV, a subsystem called
$\alpha$ cluster can easily form and often appears
in the excited states of $N=Z$ nuclei near the $\alpha$-cluster
thresholds~\cite{Ikeda68}.
Thus, it is reasonable to assume such a subsystem
as internal degrees of freedom for describing
the low-lying states of nuclear systems.
Cluster models involving $\alpha$ particles
have been applied to light nuclear systems and have succeeded
in explaining those clustering phenomena~\cite{Fujiwara80,Michel98,Cluster}.
Famous examples include the first excited $0^+$ states in $^{12}$C and $^{16}$O.
They are well explained by $\alpha+\alpha+\alpha$~\cite{Kamimura81},
and $\alpha+^{12}$C cluster models~\cite{Suzuki76a,Suzuki76b},
respectively. Such a cluster structure, e.g.,
the famous Hoyle state~\cite{Hoyle}, crucially determines
the radiative capture rate which
directly impacts on the nucleosynthesis.
Capture by heavier $N=Z$ clusters will be important
when approaching the final stage of the star burning process.

In general, the application of a fully microscopic cluster model 
(Resonating Group Method; RGM~\cite{RGM1,RGM1b,RGM2}) to
systems involving heavier clusters is much involved.
One has to derive complicated mathematical formulas,
the so-called RGM kernels, for each system.
Another problem is the effective nucleon-nucleon
interaction for the cluster model.
As exemplified in the famous
$^{12}$C and $^{16}$O problems~\cite{Itagaki95},
it is difficult to reproduce
the threshold energies in a consistent manner
for many cases.

On the other hand, a macroscopic cluster model
(Orthogonality Condition Model; OCM)
has often been used as an approximation of the RGM~\cite{OCM1,OCM2,OCM3}.
Practically, an inter-cluster potential is replaced with
a phenomenological one which reproduces properties of
the subsystem.
However, such a nucleus-nucleus potential is known to be
deep resulting in redundant bound state solutions,
which are called forbidden states,
originating from the Pauli principle
between the clusters. For example,
a phenomenological $\alpha+\alpha$ potential,
the so-called BFW potential~\cite{BFW},
allows $(n_r l)=$(00), (10), and (02) redundant forbidden states
that correspond to the total harmonic-oscillator quanta, $Q=2n_r+l<4$.
The Pauli principle between subsystems is approximately
taken into account by imposing the orthogonality conditions
to the relative wave function.
Therefore, the behavior of the wave function becomes
complicated allowing several nodes in the internal regions.

Since we have to get rid of all 
forbidden states from the solution of the many-body Schr\"odinger equation,
applying such a deep potential to many- and heavy-cluster systems
is still hard.
The removal of the forbidden states has been practically achieved, e.g., by
using the projection method~\cite{Kukulin}.
However, the method induces some numerical instability due to
a large factor, typically $\sim 10^{3-6}$,
multiplied to the projection operator.
Though the OCM have been successfully applied
  to multi-cluster systems,
the application is limited only
to a few light-cluster systems
(See, for example,
Refs.~\cite{Theeten06,Funaki08,Hiyama10,Horiuchi14}).

To extend the macroscopic model
to systems involving heavier clusters, such as $^{16}$O and $^{40}$Ca,
a shallow potential, which has no redundant bound state,
is advantageous for applications.
For example, the Ali-Bodmer potential~\cite{AB}
is a phenomenological shallow potential
which reproduces well the low-energy $\alpha+\alpha$ scattering phase shifts
although it accommodates no bound state.
Though such a shallow potential is strongly angular-momentum dependent,
its application to many-cluster systems is in general easier
than that with the OCM. See, early and recent applications
to the triple-$\alpha$ reactions~\cite{Tursunov03,Ishikawa13,Suno15}. 

We want to use such a shallow inter-cluster
potential for studying multi-cluster systems.
A supersymmetric (SUSY) transformation
offers a prescription to obtain a phase-shift-equivalent
shallow-singular potential from any deep potential having
a number of bound state solutions~\cite{Baye87,Baye04,Baye14}.
(The $r^{-2}$ singularity at the origin is needed
to improve the high energy behavior of the phase shift obtained
with the shallow potential~\cite{Swan63,Swan68}).
To remove the redundant bound states,
the transformation induces a repulsive component which pushes
the internal wave function to the outer regions to reduce the number of nodes
while keeping the phase shift invariant.
However, even though the transformed potential
is phase equivalent to the original one,
the wave functions are modified resulting in differences in the expectation 
values of observables with respect to the original potential.
See, for example, Refs.~\cite{Tursunov03,Suzuki08}
for $^{12}$C$=\alpha+\alpha+\alpha$
and Ref.~\cite{Pinilla11} for $^{6}$He$=\alpha+n+n$.

The modifications introduced by the SUSY transformation 
have not been quantified yet. 
In order to analyse them in this paper,
we investigate the behavior of the relative wave functions
of an $\alpha+$nucleus macroscopic cluster model.
We take specifically the $\alpha+^{16}$O and $\alpha+^{40}$Ca systems
towards future applications to many-body systems
such as $^{21}$Ne ($^{21}$Na)$= ^{16}$O$+\alpha+n$ $(p)$,
$^{24}{\rm Mg}= ^{16}$O$+\alpha+\alpha$,
$^{45}$Ti ($^{45}$V)$= ^{40}$Ca$+\alpha+n$ $(p)$,
and $^{48}$Cr$= ^{40}$Ca$+\alpha+\alpha$.

The paper is organized as follows:
Section~\ref{formulation.sec}
briefly explains the SUSY transformations
used in this paper to generate the shallow-singular potential
from the deep potential.
Some definitions of physical quantities
for the two-cluster model are summarized
in Sec.~\ref{defs.sec}.
Section~\ref{pot.sec} provides
phenomenological deep potentials of the
$\alpha+^{16}$O and $\alpha+^{40}$Ca systems,
and their properties are discussed
in comparison with experimental data.
In Sec.~\ref{results.sec},
we compare the relative wave functions generated from
the deep and SUSY transformed potentials,
and quantify how those differences appear
in observables such as the nuclear radius and electric-multipole
transitions and sum rules.
A summary is given in Sec.~\ref{conclusion.sec}.

\section{Formulation}
\label{formulation.sec}

\subsection{Derivation of a standard supersymmetric transformed potential}

To get a phase-equivalent supersymmetric transformed potential consistent with
the deep potential, we follow the prescription
given in Refs.~\cite{Baye87,Baye87b,Baye04}.
We consider a two-spinless-cluster system
that interacts only with a central potential.
In such a case, the relative wave function with energy $E$ 
and orbital momentum $l$ 
can be factorized into a radial part $\chi_l(E,r)/r$ 
and an angular part $Y_{lm}(\hat{\bm{r}})$.
In this section, for the sake of simplicity,
we omit the quantum numbers and radial dependence $r$
from the radial wave function, otherwise needed.
The one-dimensional radial differential equation is written as 
\begin{align}
  H\chi(E) = \left(-\frac{d^2}{dr^2}+V\right) \chi(E) = E\chi(E)
\end{align}
in units of $\hbar=2m=1$ where $m$ is
the reduced mass of the two clusters.
The central effective potential, $V$, includes the nuclear, Coulomb and
centrifugal $\frac{l(l+1)}{r^2}$ terms.

Let us assume that we have a local, regular
and deep nuclear potential which accommodates
$n$ forbidden bound-state solutions.
The ground state is eliminated by supersymmetric
transformations with a two-step procedure.
The initial Hamiltonian is factorized into two first-order operators 
$L_0^\pm = \pm d/dr + d\ln \chi(E_0)/dr$,
the so-called intertwining operators, as 
$H = L^+_0L_0^-+E_0$
where the ground-state energy $E_0$ is taken as factorization energy 
and $\chi(E_0)$ is the ground-state wave function \cite{Sukumar85}. 
The SUSY partner, $H_1$, of $H_0$ is defined by
$H_1 = L^-_0L_0^++E_0$.  
The lowest bound state is removed in $H_1$ but the phase shift is also modified. 
To recover the original phase shift, one performs another SUSY transformation
by factorizing $H_1$ in the form $H_1=L_1^+L_1^-+E_0$ 
with $L_1^\pm = \pm d/dr + d\{\ln\int_0^r dt\,[\chi(E_0)]^2/\chi(E_0)\}/dr$
and defining its partner $H_2=L_1^-L_1^++E_0$. 
The corresponding potential obtained by
those two steps can be summarized in the form \cite{Baye87,Baye87b,Baye04}
\begin{align}
  V_2=V-2\frac{d^2}{dr^2}\ln\int_0^r dt\,[\chi(E_0)]^2.
\label{V2}
\end{align}
It should be noted that the potential behaves
as $\sim(l+2)(l+3)/r^2$ around the origin
which ensures to satisfy a generalized Levinson theorem~\cite{Swan63}.
The SUSY partners have identical spectra
except for the number of bound states.
In other words, the two potentials obtained
by this method provide exactly the same phase shifts.
The physical wave function
of each Hamiltonian is related to the other one by the intertwining operators.

The above procedure will be repeated
until all the unphysical bound states are removed.
Since the integral in Eq.~(\ref{V2})
does not vanish and can thus not lead to singularities 
at finite distance in the transformed potential,
the factorization energy $E_0^{(i)}$ $(i>1)$ can be the energy
of any excited state to be suppressed.
The final form of the SUSY transformed potential reads
\begin{align}
  V_{\rm SUSY}=V_{2n}=V-2\sum_{i=0}^{n-1}
  \frac{d^2}{dr^2}\ln\int_0^r dt\,[\chi(E_0^{(i)})]^2,
\end{align}
where $n$ denotes the number of redundant or forbidden bound states.
A number $n$ of removal manipulations induces
the additional $\sim(2n+l)(2n+l+1)/r^2$ $(r\to 0)$
singularity~\cite{Baye04,Baye14}.
The potential becomes `more singular' as
the singularity parameter $2n+l$ increases with larger $n$.

\subsection{``Designed'' supersymmetric transformed potential}
\label{susym.sec}

In the previous subsection,
we have discussed how we obtain the phase-equivalent potential
by eliminating the forbidden states by the SUSY transformation.
Here we generate another supersymmetric transformed potential.
In the SUSY prescription,
we can remove and reintroduce any bound state of the spectrum~\cite{Baye87b}.
The transformation allows an arbitrary free parameter
that may be fixed to reproduce one physical quantity.
The resultant wave function is phase-equivalent
but it is modified depending on
the choice of the arbitrary parameter.
With this procedure, we can ``design'' the SUSY potential
so as to reproduce a physical quantity other then the phase shift.

After elimination of all the redundant bound states, 
the physical bound state at energy $E_0^{(n)}$ with wave function
$\chi(E_0^{(n)})$ is removed by a factorization similar to the factorization of the initial $H$ 
described above and the non-equivalent Hamiltonian $H_{2n+1}$ is obtained. 
Then $H_{2n+1}$ is factorized with the factorization energy $E_0^{(n)}$, 
$H_{2n+1}=L_{2n+1}^+L_{2n+1}^-+E_0^{(n)}$,
with $L^{\pm}_{2n+1}=\pm d/dr + d\ln \varphi(E_0^{(n)})/dr$, where 
\begin{align}
\varphi(E_0^{(n)})=
\frac{1}{\chi(E_0^{(n)})}\left(\beta+\int_r^\infty dt\,[\chi(E_0^{(n)})]^2\right)
\label{wfb.eq}
\end{align}
is an unbound solution both at the origin and at infinity.      
Its SUSY partner $H_{2n+2}=L_{2n+1}^-L_{2n+1}^++E_0^{(n)}$ has again a ground state 
at energy $E_0^{(n)}$ and the phase equivalence to $H_{2n}$ and thus to $H$ is restored. 
Now an arbitrary parameter $\beta$ is introduced which modifies
the properties of the ground-state wave function.
In Eq.~(\ref{wfb.eq}), since the wave function $\chi(E_0^{(n)})$
is normalized,
$\beta$ takes any value outside $[-1,0]$. Indeed, the integral
takes its values within $[0,1]$.
For $\beta=-1$, the state is removed like in the previous subsection; 
  for $\beta \in [-1,0]$,
the potential is singular at finite distance.
The new wave function of the ground state is
\begin{align}
  \chi_{2n+2}(E_0^{(n)})=\sqrt{\beta(\beta+1)}\, \chi(E_0^{(n)})
  \left(\beta+\int_r^\infty dt\,[\chi(E_0^{(n)})]^2\right)^{-1}.
\end{align}
Its asymptotic form $(r\to \infty)$ is
\begin{align}
  \chi_{2n+2}(E_0^{(n)})\to \sqrt{(\beta+1)/\beta}\, \chi(E_0^{(n)}).
\label{asym.eq}
\end{align}
Hence the initial asymptotic normalization constant is multiplied
by $\sqrt{(\beta+1)/\beta}$.
The new phase-equivalent potential (SUSY-$\beta$) can be calculated by
\begin{align}
  V_{{\rm SUSY-}\beta}=V_{2n}-2\frac{d^2}{dr^2}
  \ln\left|\beta+\int_r^\infty dt\,[\chi(E_0^{(n)})]^2\right|.
\end{align}
The singularity of $V_{{\rm SUSY-}\beta}$ is unchanged
as $(2n+l)(2n+l+1)/r^2$ $(r\to 0)$. For practical purposes,
the $\beta$ value is fixed so as to reproduce
the desired physical quantity
which will be discussed later in Sec.~\ref{phys.sec}.

\section{Physical quantities in the two-cluster systems}
\label{defs.sec}

In this section, we summarize definitions of the physical quantities
used in the $\alpha$-nucleus systems.
The root-mean-square (rms) radius composed of a two-cluster, $C_1 + C_2$, system
with non-integer mass number
$A$ can be evaluated by
\begin{align}
  r(C_1+C_2)=
\sqrt{\frac{A_1[r(C_1)]^2+A_2[r(C_2)]^2}{A_1+A_2}+\frac{A_1A_2}{(A_1+A_2)^2}\left<r^2\right>},
\end{align}
where $A_1$ $(A_2)$, $r(C_1)$ $(r(C_2))$,
and $\left<r^2\right>$ are the non-integer mass number
and the rms radius of cluster 1 (2), and the mean-square distance
between the two clusters.

We will also evaluate the
reduced electric-multipole ($E\lambda$) transition probability
with multipolarity $\lambda$ defined as
\begin{align}
  B(E\lambda;J_i^{\pi_i} \to J_f^{\pi_f})=
  \frac{1}{2J_i+1}\sum_{M_f,M_i,\mu}\left|\right<J_f^{\pi_f}M_f|\mathcal{M}_{\lambda\mu}
  |J_i^{\pi_i}M_i\left>\right|^2.
\end{align}
We remark that the initial parity $\pi_i$ changes to
$\pi_f$ with $(-1)^{\lambda}$.
By ignoring the internal excitation of the two clusters,
effective $E\lambda$ operators,
which only act on the relative wave function between the two clusters,
are given by (see, for example, Appendix B of Ref.~\cite{Horiuchi14b})
\begin{align}
  \mathcal{M}_{\lambda\mu}&=
  \frac{Z_1A_2^\lambda+Z_2(-A_1)^\lambda}{(A_1+A_2)^\lambda}
er^{\lambda}Y_{\lambda\mu}(\bm{\hat{r}}),
\end{align}
    where $Z_1$ ($Z_2$) is the charge number of the cluster 1 (2).
    We remark that
    the electric-dipole ($\lambda=1$) operator vanishes
    when a system consists
    of two $N=Z$ clusters with integer mass numbers.
Though it does not vanish with the non-integer mass numbers,
we however do not calculate the $E1$ transition probabilities because
the formula has no physical basis in this case~\cite{Baye18}. 

\section{Phenomenological $\alpha$+$^{16}$O and $\alpha$+$^{40}$Ca potentials}
\label{pot.sec}

\subsection{$\alpha$-nucleus potentials}

The phenomenological $\alpha$-nucleus potentials are assumed
to be a parity-dependent-single Gaussian form as
\begin{align}
V(r)=(V_0+V_r\hat{P}_r)\exp(-\mu_r r^2),
\end{align}
where $\hat{P}_r$ is the parity operator
that changes $\bm{r}$ into $-\bm{r}$.
This form factor was successful in $\alpha$+$\alpha$~\cite{BFW}
and $\alpha$+$^{16}$O~\cite{Kruppa90} scattering problems.
In addition to the nuclear potential we include
the Coulomb term as
\begin{align}
  V_C(r)=2Z_2\frac{e^2}{r}{\rm erf}(\nu r)
\end{align}
with the error function \cite{Abramowitz}
where $\nu$ is fixed so as to reproduce
the potential value of a uniform charge distribution
with a sphere radius $R_C=\sqrt{5/3}[r_C(C_1)+r_C(C_2)]$ at the origin,
where the observed charge radii employed are~\cite{Angeli13}
$r_C(^4{\rm He})=1.676$\,fm,
$r_C(^{16}{\rm O})=2.699$\,fm,
and $r_C(^{40}{\rm Ca})=3.478$\,fm.
The $\nu$ values are 0.23536 and 0.19979\,fm$^{-1}$
for the $\alpha+^{16}$O and $\alpha+^{40}$Ca systems, respectively.

We fix a set of parameters, $V_0$, $V_r$, and $\mu_r$,
so as to reproduce the low-lying observables
of $^{20}$Ne and $^{44}$Ti
for the $\alpha+^{16}$O and $\alpha+^{40}$Ca systems, respectively.
The masses of the clusters used in this paper are
$3727.3794$\,MeV, $14895.0826$\,MeV, and $37214.7160$\,MeV
for $^{4}$He, $^{16}$O, and $^{40}$Ca, respectively
~\cite{Wang17a,Wang17b}.
They correspond
to $A_\alpha =4.0015$, $A_{\rm O} =15.9905$, and $A_{\rm Ca} =39.9516$.
The potential strengths should be deep enough
to accommodate a number of forbidden bound states
with the total harmonic oscillator quanta of the cluster relative
motion, $Q = 2n_r+l < 8$ and 12
for the $\alpha+^{16}$O and $\alpha+^{40}$Ca systems, respectively.
The three independent parameters are fixed so as to reproduce
three low-lying observables among
the binding energies of the bound states
$E(J^\pi)$ and the rms radius.
We generate three sets of potentials with the following
choices of the low-lying observables:
Set A reproduces $E(0^+)$, $E(2^+)$, and $E(1^-)$;
Set B reproduces $E(0^+)$, $E(4^+)$ for $^{20}$Ne
($E(6^+)$ for $^{44}$Ti), and $E(1^-)$;
and Set C reproduces $E(0^+)$
and the rms radius of the $0^+$ ground state $r(0^+)$, and $E(1^-)$.

Table~\ref{potentials.tab} lists the potential parameters of Sets A, B and C
for the $\alpha+^{16}$O and $\alpha+^{40}$Ca systems.
The behavior of the potential form factors
is summarized as follows:
Set A shows the deepest potential at the origin
and narrowest interaction range.
Set C gives a longer interaction range than that of Set A.
Set B is intermediate between Sets A and C for $\alpha+^{16}$O
and similar to Set C for $\alpha+^{40}$Ca.
The small parity dependence of the potential
implies an inversion doublet
in a well-developed asymmetric cluster structure~\cite{Horiuchi68}.
For the $\alpha+^{40}$Ca system,
the potential is in general deeper than that of the $\alpha+^{16}$O system
because more redundant bound states are required to satisfy $Q=2n_r+l<12$.
We however note that the condition cannot be satisfied with $l=11$.
No redundant bound state with $l=11$ is found in those parameter sets.

\subsection{Energy spectrum of $^{20}$Ne and $^{44}$Ti}
  
\begin{table}[!h]
  \caption{Parameters of the $\alpha+^{16}$O
    and $\alpha+^{40}$Ca potentials and choices of fitted observables.
		See text for details.}
\label{potentials.tab}
\centering
\begin{tabular}{cccrrl}
\hline
Nucleus &&$V_0$ (MeV)&$V_r$ (MeV)&$\mu_r$ (fm$^{-2}$)&\\ 
\hline
$^{16}$O &Set A&$-$215.0445&5.9455&0.16433&$E(0^+),E(2^+),E(1^-)$\\
         &Set B&$-$169.2645&2.2725&0.12501&$E(0^+),E(4^+),E(1^-)$\\
         &Set C&$-$133.7395&0.5850&0.10641&$E(0^+),r(0^+),E(1^-)$\\
$^{40}$Ca&Set A&$-$293.0125&5.1875&0.12201&$E(0^+),E(2^+),E(1^-)$\\
         &Set B&$-$167.0665&$-$1.4805&0.062002&$E(0^+),E(6^+),E(1^-)$\\
         &Set C&$-$192.4970&$-$0.2150&0.073787&$E(0^+),r(0^+),E(1^-)$\\
\hline
\end{tabular}
\end{table}

\begin{figure}[!ht]
\centering\includegraphics[width=\linewidth]{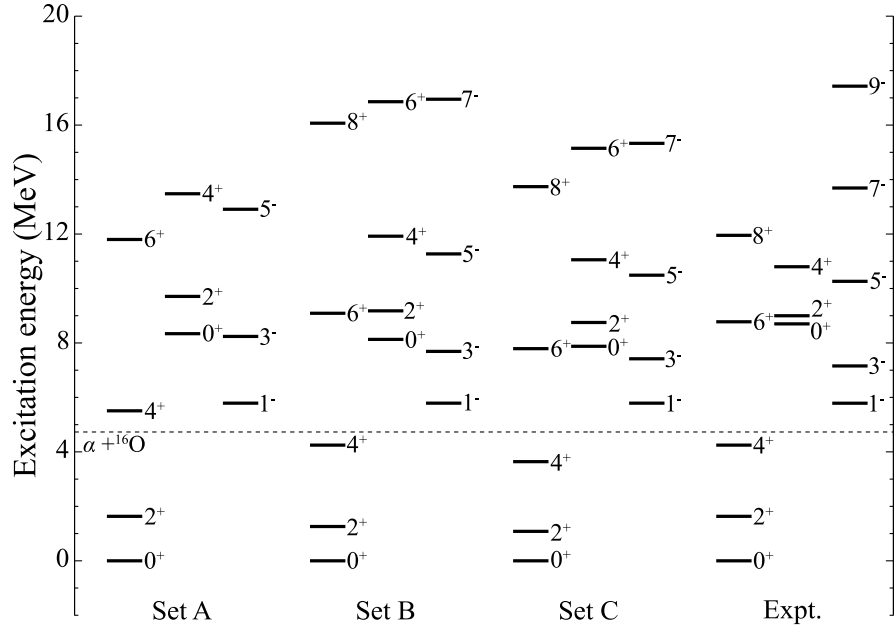}
  \caption{Energy spectrum of $^{20}$Ne with three sets of
    $\alpha+^{16}$O potentials.
    Experimental data are taken from Ref.~\cite{Tilley98}.}
\label{spect20Ne.fig}
\end{figure}

The calculated energy spectrum of $^{20}$Ne is plotted
in Fig.~\ref{spect20Ne.fig} together with the experimental data.
The $J^\pi=0^+, 2^+,$ and $4^+$ states are observed
as bound states, and the other states including
  parity inverted states~\cite{Horiuchi68} are observed as resonant states.
  These levels are well recognized as
  the rotational structure of $\alpha+^{16}$O~\cite{Horiuchi68,Kruppa90}.
  The positive-parity higher nodal excited states of the ground-state band
  are also observed at $\gtrsim 8$\,MeV.
  Here, the resonance energy $E_R$ is defined as an energy where
  the phase shift, $\delta_J$ crosses $\pi/2$.
  For a state with a broad width which does not reach $\delta_J=\pi/2$,
  we take a peak of the first derivative
  of the phase shift, $d\delta_J/dE$, as $E_R$.
In Set A, where the $0^+$ and $2^+$ energies are used to fix the potential
parameters, the $4^+$ state appears to be unbound
and the other rotational levels tend to be at higher energy
compared to the experimental data.
Sets B and C show reasonable agreement
with the observed spectrum for both the positive-
and negative-parity states.
The $0^+, 2^+$, and $4^+$ higher nodal excited states are also
in fair agreement with the experimental data.

Table~\ref{energy20Ne.tab} summarizes
the binding energies and decay widths
of $^{20}$Ne with the three sets of potential parameters.
The decay widths are evaluated with $2/(d\delta_J/dE)|_{E=E_R}$.
The observed decay widths are overestimated,
with the best agreement for Set C.
For the higher nodal states $J^\pi$,
the calculated decay widths also 
tend to be larger than the experimental data.
The fact is also found
in the early $\alpha+^{16}$O OCM calculation~\cite{Kruppa90},
whose potential parameter is intermediate between Sets B and C.

\begin{table}[!ht]
  \caption{Energies and decay widths of $^{20}$Ne with
    the $\alpha+^{16}$O model in MeV units.
    The energies are measured from the $\alpha+^{16}$O threshold.
    The lower rows list those of the nodal excited states
      of the ground-state band.
  Experimental data are taken from Ref.~\cite{Tilley98}.}
\label{energy20Ne.tab}
\centering
\begin{tabular}{cccccccccccc}
\hline
&\multicolumn{2}{c}{Set A}&&\multicolumn{2}{c}{Set B}&&\multicolumn{2}{c}{Set C}&&\multicolumn{2}{c}{Expt.}\\
\cline{2-3}\cline{5-6}\cline{8-9}\cline{11-12}
$J^\pi$&$E$ &$\Gamma$&& $E$&$\Gamma$&& $E$&$\Gamma$&&$E$&$\Gamma$\\
\hline
$0^+$ &$-$4.730&--&&$-$4.730&--&&$-$4.730&--&&$-$4.730&--\\ 
$2^+$ &$-$3.096&--&&$-$3.470&--&&$-$3.649&--&&$-$3.096&--\\ 
$4^+$ &0.779 &$<$3$\times 10^{-10}$&&$-$0.482&--&&$-$1.088&--&&$-$0.482&--\\ 
$6^+$ &7.07&2.0$\times 10^{-2}$&&4.36 &1.0$\times 10^{-3}$&&3.06&6.2$\times 10^{-5}$&&4.048&1.1$\times 10^{-4}$\\ 
$8^+$ &16.22&0.17&&11.34&3.5$\times 10^{-2}$&&9.01&1.0$\times 10^{-2}$&&7.22&3.5$\times 10^{-5}$\\ 
$1^-$ &1.058&1.6$\times 10^{-5}$&&1.058&3.4$\times 10^{-5}$&&1.058&5.5$\times 10^{-5}$
&&1.058&2.8$\times 10^{-5}$\\ 
$3^-$ &3.51&8.0$\times 10^{-2}$&&2.96&5.0$\times 10^{-2}$&&2.69&4.0$\times 10^{-2}$
&&2.426&8.2$\times 10^{-3}$\\ 
$5^-$ &8.18&0.76&&6.54&0.48&& 5.76& 0.38&&
5.532&0.145\\
$7^-$ &15.61&1.84&&12.22&1.22&& 10.60& 0.96&&8.962&0.31\\ 
$9^-$ &26.31&2.36&&20.38&1.59&& 17.57& 1.27&&12.70&0.22\\ 
\hline
$0^+$&3.61&16.9&&3.40&5.42&& 3.13& 2.90&&$\approx4.0$&$>0.8$\\
$2^+$&4.98&18.6&&4.45&6.76&& 4.02& 3.84&&4.27&$\approx0.8$ \\
$4^+$&8.75&21.0&&7.19&9.05&& 6.32& 5.60&&6.07&0.35\\
\hline
\end{tabular}
\end{table}

\begin{table}[!ht]
  \caption{Point-matter rms radii
    and electric-quadrupole ($E2$) reduced transition probabilities
    of $^{20}$Ne with the $\alpha+^{16}$O models.
    Three sets of potential parameters are employed.
    The point-proton rms radii of $^{4}$He and $^{16}$O are 
    1.455 and 2.568\,fm, respectively, extracted from Ref.~\cite{Angeli13}.
    Values in parentheses are in units of W.u.$(=
    \frac{(1.2)^{2\lambda}}{4\pi}(\frac{3}{\lambda+3})^2A^{2\lambda/3}
    \ (e^2{\rm fm}^{2\lambda}$)~\cite{BM}).
    The experimental $B(E2)$ value is taken from Ref.~\cite{Tilley98}.
    }
\label{rms20Ne.tab}
\centering
\begin{tabular}{ccccccc}
\hline
&&Set A&Set B&Set C&Expt.\\
\hline
$r(0^+)$ (fm) &&2.768  &2.840& 2.889& 2.889$\pm$0.002\\ 
$r(2^+)$ (fm) &&2.766  &2.837& 2.884\\ 
$r(4^+)$ (fm) &&unbound&2.827& 2.872\\ 
$B(E2;2^+\to 0^+)$ ($e^2$fm$^4$)
&& 30.2 (9.4) & 43.9 (13.6) & 54.7 (17.0) &65.5$\pm 3.2$ (20.3$\pm 1.0$)\\
$B(E2;4^+\to 2^+)$ ($e^2$fm$^4$)
&&--          & 58.4 (18.1) & 72.2 (22.4) &71$\pm 6$ (22$\pm 2$)\\
\hline
\end{tabular}
\end{table}

Table~\ref{rms20Ne.tab} lists rms radii and
reduced electric-quadrupole ($E2$) transition probabilities.
The experimental data are also listed for comparison.
The rms radius decreases with increasing
angular momentum because
the attractive pocket of the potential
becomes narrower with increasing
centrifugal barrier.
Set C reproduces
the $B(E2)$ value best among the three sets.
Remarking that Set C is determined in such a way
so as to reproduce the binding energy
and the rms radius of the ground state of $^{20}$Ne,
the size of the ground-state wave function
is essential to get a reasonable moment
of inertia for describing the $\alpha+^{16}$O 
rotational structure in $^{20}$Ne.

\begin{figure}[!ht]
  \centering\includegraphics[width=\linewidth]{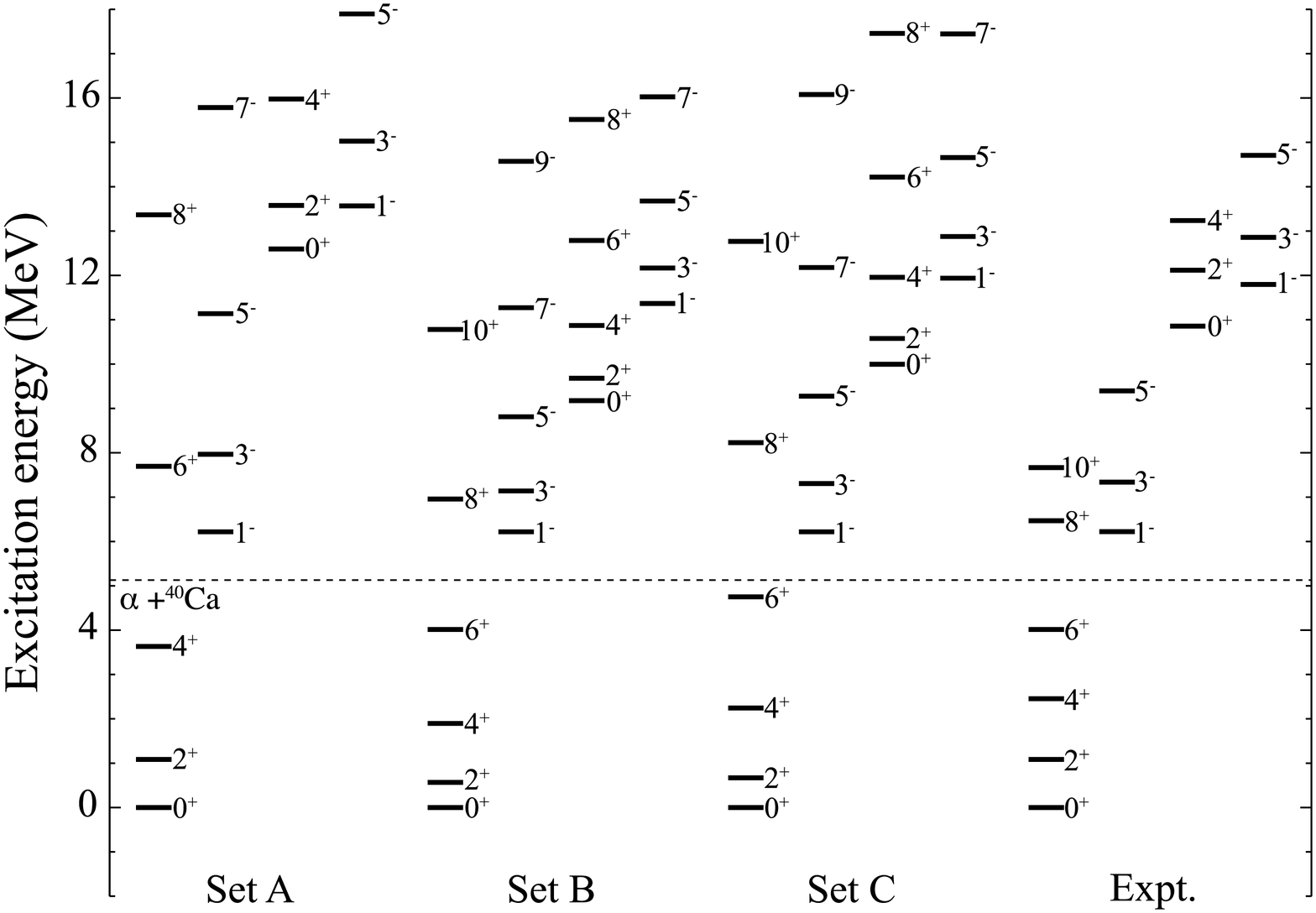}
  \caption{Energy spectrum of $^{44}$Ti with three sets of
    $\alpha+^{40}$Ca potentials.
    Experimental data are taken from Ref.~\cite{Chen11}.}
\label{spect44Ti.fig}
\end{figure}

\begin{table}[!ht]
  \caption{Same as Table~\ref{rms20Ne.tab}.
    but for $^{44}$Ti with the $\alpha+^{40}$Ca model.
    The point-proton rms radius of $^{40}$Ca is
    3.377\,fm extracted from Ref.~\cite{Angeli13}.
      The experimental $B(E2)$ value is taken from Ref.~\cite{Chen11}.}
\label{rms44Ti.tab}
\centering
\begin{tabular}{ccccccc}
\hline
&&Set A&Set B&Set C&Expt.\\
\hline
$r(0^+)$ (fm)&&3.437  &3.548 & 3.515&3.515$\pm$0.005\\ 
$r(2^+)$ (fm)&&3.436  &3.547 & 3.513&\\ 
$r(4^+)$ (fm)&&3.434  &3.542 & 3.509&\\ 
$r(6^+)$ (fm)&&unbound&3.535 & 3.503&\\ 
$B(E2;2^+\to 0^+)$ ($e^2$fm$^4$)&&59.9 (6.5)&156.9 (17.0) &122.4 (13.3)&120$\pm 30$ (13$\pm 3$)\\ 
$B(E2;4^+\to 2^+)$ ($e^2$fm$^4$)&&82.6 (8.9)&214.3 (23.2) &167.5 (18.1)&280$\pm 60$ (30$\pm 6$)\\ 
$B(E2;6^+\to 4^+)$ ($e^2$fm$^4$)&&--        &215.8 (23.4) &169.2 (18.3)&160$\pm 30$ (17$\pm 3$)\\ 
\hline
\end{tabular}
\end{table}

\begin{table}[!ht]
  \caption{Same as Table~\ref{energy20Ne.tab} but for $^{44}$Ti with the $\alpha+^{40}$Ca model. Experimental data are taken from Ref.~\cite{Chen11}.}
\label{energy44Ti.tab}
\centering
\begin{tabular}{ccccccccccc}
\hline
     &\multicolumn{2}{c}{Set A}&&\multicolumn{2}{c}{Set B}&&\multicolumn{2}{c}{Set C}&&Expt.\\
    \cline{2-3}\cline{5-6}\cline{8-9}\cline{11-11}
$J^\pi$ &$E$ &$\Gamma$&& $E$&$\Gamma$&& $E$&$\Gamma$&&$E$\\
    \hline
    $0^+$ &$-$5.127&--&&$-$5.127&--&&$-$5.127&--&&$-$5.127\\ 
    $2^+$ &$-$4.044&--&&$-$4.560&--&&$-$4.457&--&&$-$4.044\\ 
    $4^+$ &$-$1.496&--&&$-$3.229&--&&$-$2.883&--&&$-$2.673\\ 
    $6^+$ &2.57&$<$5$\times 10^{-11}$&&$-$1.112&--&&$-$0.377&--&&$-$1.112\\ 
    $8^+$ &8.24 &4.5$\times 10^{-5}$ &&1.83&$<$2$\times 10^{-10}$&&3.11 &$<$1$\times 10^{-10}$&&1.34 \\ 
   $10^+$ &15.68&3.7$\times 10^{-3}$ &&5.65&6.3$\times 10^{-8}$  &&7.65 &3.3$\times 10^{-6}$  &&2.54 \\ 
    $1^-$ &1.093&$<$1$\times 10^{-12}$&&1.093&$<$1$\times 10^{-12}$&&1.093&$<$1$\times 10^{-12}$&&1.093\\ 
    $3^-$ &2.84&2.9$\times 10^{-8}$  &&2.01&$\approx$5$\times 10^{-10}$ &&2.18 &1.2$\times 10^{-9}$  &&2.21 \\ 
    $5^-$ &6.01&6.2$\times 10^{-4}$  &&3.69&1.1$\times 10^{-5}$  &&4.15 &3.2$\times 10^{-5}$  &&4.27 \\
    $7^-$ &10.66&3.4$\times 10^{-2}$ &&6.14&1.5$\times 10^{-3}$  &&7.05 &3.5$\times 10^{-3}$  &&--\\ 
    $9^-$ &16.93&0.22                &&9.45&1.8$\times 10^{-2}$  &&10.95&3.5$\times 10^{-2}$  &&--\\ 
    \hline
    $0^+$ &7.47  &1.20  &&4.05  &1.2$\times 10^{-2}$ &&4.87  &6.9$\times 10^{-2}$ &&5.73 \\
    $2^+$ &8.45  &1.70  &&4.55  &2.6$\times 10^{-2}$ &&5.45  &0.12                &&6.99 \\
    $4^+$ &10.85 &3.03  &&5.74  &9.0$\times 10^{-2}$ &&6.83  &0.29                &&8.11 \\
    $1^-$ &8.44  &8.22  &&6.24  &2.48                &&6.81  &3.38                &&6.67  \\
    $3^-$ &9.90  &10.2  &&7.04  &2.99                &&7.75  &4.08                &&7.73  \\
    $5^-$ &12.77 &13.2  &&8.55  &3.84                &&9.53  &5.21                &&9.58  \\
 \hline
\end{tabular}
\end{table}

Figure~\ref{spect44Ti.fig} displays the spectrum of $^{44}$Ti.
Similarly to the $\alpha+^{16}$O case,
Set A produces a large moment of inertia resulting in
too large an energy splitting between the intraband states,
and Sets B and C fairly well reproduce 
the $\alpha+^{40}$Ca rotational structure~\cite{Michel98,Kimura06}, that is,
the low-lying positive- and negative-parity levels
as well as their nodal excited states observed at $\gtrsim 12$\,MeV,
and the $B(E2)$ values as shown in Table~\ref{rms44Ti.tab}.
We also list, in Table~\ref{energy44Ti.tab}, the binding energies and
decay widths of $^{44}$Ti with the $\alpha+^{40}$Ca model,
although no $\alpha$-decay
width is observed in such low-lying states.

\section{Tests of the SUSY transformed potentials}
\label{results.sec}

Hereafter we employ Set C as
the initial $\alpha$-nucleus deep potentials
to be transformed by the SUSY prescription.

\subsection{Physical properties of $\alpha+$nucleus systems}
\label{phys.sec}

\begin{table}[!ht]
  \caption{Root-mean-square distances between the clusters, electric-quadrupole ($E2$) and hexadecapole ($E4$)
    reduced transition probabilities between the bound states.
    Both the results of $^{20}$Ne and $^{44}$Ti are listed.}
\label{phys20Ne.tab}
\centering
\begin{tabular}{cccccccc}
\hline
$^{20}$Ne&&\multicolumn{1}{c}{SUSY}&&\multicolumn{1}{c}{SUSY-$\beta$}&&\multicolumn{1}{c}{OCM}\\
\hline
$\sqrt{\left<r^2\right>}(0^+)$ (fm)&& 4.391&& 4.067&& 4.067  \\
$\sqrt{\left<r^2\right>}(2^+)$ (fm)&& 4.353&& 4.027&& 4.046  \\
$\sqrt{\left<r^2\right>}(4^+)$ (fm)&& 4.254&& 3.924&& 3.991  \\
$B(E2;2^+\to 0^+)$ (W.u.) && 23.1 && 16.9 && 17.0 \\
$B(E2;4^+\to 2^+)$ (W.u.) && 30.8 && 22.4 && 22.4 \\
$B(E4;4^+\to 0^+)$ (W.u.) && 47.3 && 25.5 && 37.2 \\
$B(E4;4^+\to 2^+)$ (W.u.) && 59.7 && 32.1 && 47.7 \\
\hline
$^{44}$Ti&&\multicolumn{1}{c}{SUSY}&&\multicolumn{1}{c}{SUSY-$\beta$}&&\multicolumn{1}{c}{OCM}\\
\hline
$\sqrt{\left<r^2\right>}(0^+)$ (fm)&&5.157&&4.657&&4.657  \\
$\sqrt{\left<r^2\right>}(2^+)$ (fm)&&5.132&&4.631&&4.642  \\
$\sqrt{\left<r^2\right>}(4^+)$ (fm)&&5.071&&4.569&&4.607  \\
$\sqrt{\left<r^2\right>}(6^+)$ (fm)&&4.967&&4.463&&4.547  \\
$B(E2;2^+\to 0^+)$ (W.u.) &&20.0 &&13.3 &&13.3 \\
$B(E2;4^+\to 2^+)$ (W.u.) &&27.5 &&18.2 &&18.1 \\
$B(E2;6^+\to 4^+)$ (W.u.) &&28.3 &&18.5 &&18.3 \\
$B(E4;4^+\to 0^+)$ (W.u.) &&53.6 &&23.7 &&36.7 \\
$B(E4;4^+\to 2^+)$ (W.u.) &&68.4 &&30.3 &&47.4 \\
$B(E4;6^+\to 2^+)$ (W.u.) &&76.0 &&33.2 &&51.1 \\
$B(E4;6^+\to 4^+)$ (W.u.) &&58.4 &&25.4 &&40.7 \\
\hline
\end{tabular}
\end{table}

Table~\ref{phys20Ne.tab} lists the rms distances between the clusters   
of the bound states with spin-parity $J^\pi=0^+$, $2^+$, 4$^+$,
reduced $E2$ and electric-hexadecapole ($E4$) transition
probabilities of $^{20}$Ne with the $\alpha+^{16}$O model.
The relative wave functions between the clusters are generated
by the SUSY transformed potentials as well as
the initial deep OCM potential.
Despite the fact that the SUSY and deep potentials
give exactly the same phase shift,
the rms distances and the $B(E2)$ values
are actually modified and increased by
the standard SUSY transformation.
Those observables are somewhat sensitive
to the wave function at short distances.
On the contrary, the changes in the $B(E4)$ values
are relatively smaller
than that of the $B(E2)$ case. Since the operator
is proportional to $r^4$,
the $B(E4)$ values are less sensitive
to the internal regions of the wave function,
whereas the external regions contribute largely
to the $B(E4)$ matrix element.
Recalling that we fix the parameters of the deep potential
in such a way so as to reproduce the binding energy and rms radius
of the ground-state wave function of $^{20}$Ne,
the standard SUSY transformation is no longer a desired potential.

Therefore, we consider another SUSY transformation
(SUSY-$\beta$) prescribed in Sec.~\ref{susym.sec}
with the arbitrary parameter
$\beta$ being
fixed so as to reproduce the rms radius of the ground state.
The choice of the $\beta$ value can be angular-momentum dependent.
We, however, transform all the $J^\pi$ states having
the physical bound states
with the same $\beta$ value for the sake of simplicity.
The $\beta$ values are  $-1.8670$
and $-1.3791$ for $^{20}$Ne and $^{44}$Ti,
respectively. 
These values are negative in order to provide a narrower pocket in the potential.
As shown in Table~\ref{phys20Ne.tab},
the $B(E2)$ values and rms radii of the bound $2^+$ and $4^+$ states
are remedied by this transformation.
However, the $B(E4)$ value shows some deviation from
that with the OCM calculation. We will return 
to this matter later in Sec.~\ref{trdens.sec}.

The same holds for $^{44}$Ti.
Table~\ref{phys20Ne.tab} also lists rms distances between the clusters,
reduced $E2$, $E4$
transition probabilities of $^{44}$Ti in the $\alpha+^{40}$Ca potential models.
As already seen in the $^{20}$Ne case,
the rms distances and $B(E\lambda)$ transition probabilities
are modified by the standard SUSY transformation.
The modification becomes somewhat moderate
for higher multipoles due to the $r^{\lambda}$ factor
in the electric-multipole operator with a rank $\lambda$.
In fact, similarly to the $^{20}$Ne case,
  the SUSY-$\beta$ transformation gives smaller $B(E4)$ values.
Though one always needs to check how observables of interest
are modified by the transformation,
as long as the rms radius and $B(E2)$ values are of interest,
the SUSY-$\beta$ seems to provide a reasonable
forbidden-state-free potential which can be used
for studying more than three-cluster systems.

\subsection{Supersymmetric transformed potentials
and wave functions}

\begin{figure}[!ht]
\centering\includegraphics[width=\linewidth]{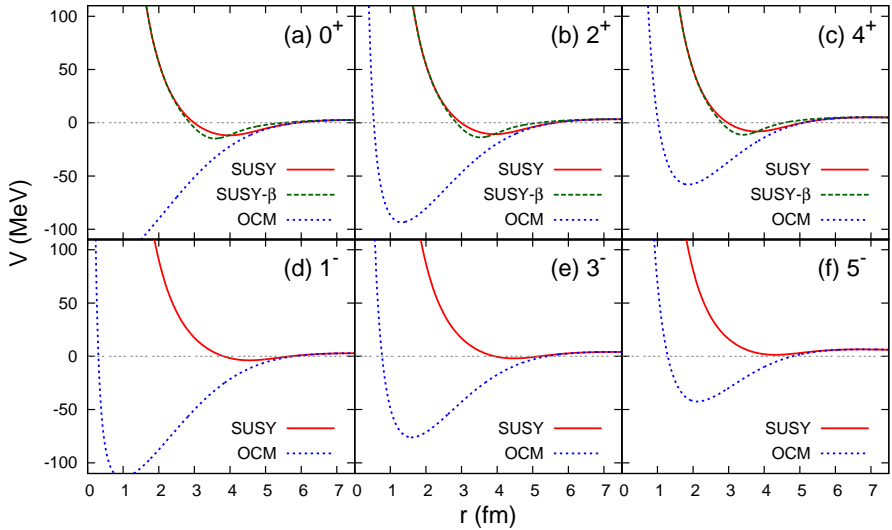}
\caption{Comparison of the SUSY transformed and original deep
    (OCM) potentials of the $\alpha+^{16}$O system for the 
  $J^\pi=$ (a) 0$^+$, (b) 2$^+$, (c) 4$^+$, (d) 1$^-$, (e) 3$^-$, and (f) 5$^-$ states.
  The Coulomb and centrifugal potentials
  are included in the plot.
Set C potential is employed.}
\label{pot20Ne.fig}
\end{figure}

Figure~\ref{pot20Ne.fig} plots
the SUSY transformed $\alpha+^{16}$O potentials
as well as the original deep potential for different $J^\pi$ states.
The Coulomb and centrifugal potentials are included
in the plots. No SUSY-$\beta$ potential exists for
the negative parity states
as they have no physical bound state.

The SUSY transformed potentials at $r\lesssim 5$\,fm
behave quite differently from the original deep potential.
For all the $J^\pi$ states, including the $0^+$ state,
we see a strong repulsion for $r\lesssim 3$\,fm generated
from the SUSY transformations
pushing the internal wave function to outer regions.
This behavior is a general consequence of removing
the redundant bound states while keeping the phase shift invariant. 
We see attractive pockets
at $\sim 4$\,fm for the $0^+$, $2^+$, and $4^+$ states
which allow us to obtain one physical bound-state solution.
The attractive pocket is somewhat shifted to inner regions with
the SUSY-$\beta$ transformation with a negative $\beta$ value because
this transformation always gives a smaller rms radius than that
of the standard SUSY transformation
shown in Table~\ref{phys20Ne.tab}.

For the negative-parity partial waves,
the pocket is not deep enough to accommodate any bound state.
The lowest state appears as a resonance with a narrow width.
We again remark that all the potentials give
exactly the same phase shift.

\begin{figure}[!ht]
\centering\includegraphics[width=\linewidth]{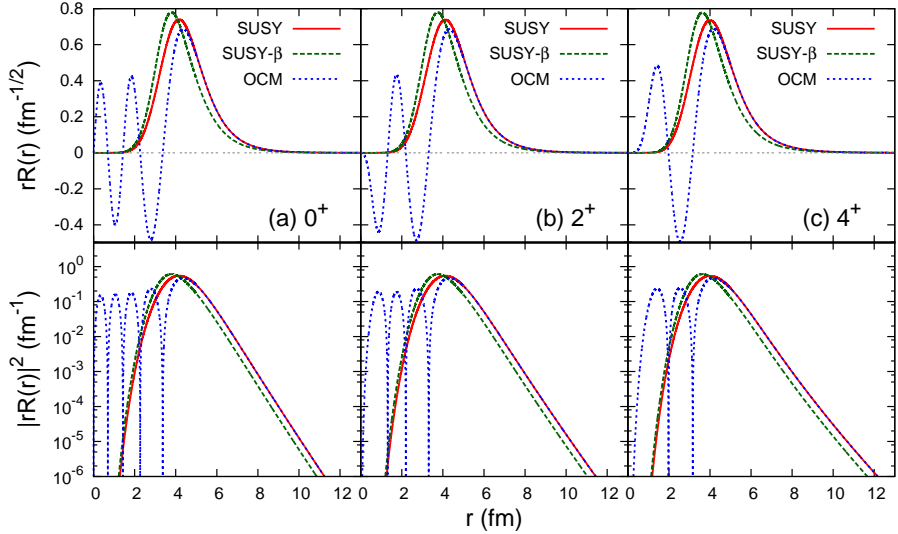}
  \caption{Comparison of  
    the bound-state wave functions 
  of the lowest (a) 0$^+$, (b) $2^+$, and (c) $4^+$ physical states of $^{20}$Ne
  obtained with the SUSY transformed and the original deep (OCM)
potentials.}
\label{wave20Ne.fig}
\end{figure}

We solve the Schr\"{o}dinger equation numerically
for each $J$ with those potentials.
The second-order differential equation is precisely solved by
a finite-difference method, the Numerov method,
with the boundary condition that the logarithmic derivative
of the numerical wave function matches the Whittaker function~\cite{Whittaker,Abramowitz} at large distances.
Figure~\ref{wave20Ne.fig} plots
the radial wave function of physical bound states of $^{20}$Ne
generated by the SUSY transformed
and original deep potentials.
Logarithmic plots of the probability distribution
are also presented in order to clearly
see the asymptotics of the wave functions.
When the OCM calculation is made,
all the displayed physical wave functions obtained with the deep potential
exhibit several nodes that satisfy
the total harmonic-oscillator quanta $Q=2n_r+l=8$
and the internal wave function $r\lesssim 4$\,fm
changes drastically with the angular momentum increases.
In contrast, the SUSY transformed potentials provide
nodeless wave functions for these physical bound states,
and the behavior of the wave functions in the internal regions
does not depend so much on the angular momentum.
As seen in Fig.~\ref{wave20Ne.fig}, the wave functions obtained by
the original deep and standard SUSY transformed potentials
are identical for $r \gtrsim 5$\,fm, which can be expected
from the phase-equivalence.
The SUSY-$\beta$ potential also gives a nodeless wave function
but the peak position is a little shifted
to the inner regions corresponding to the position
of the attracted potential pocket as already seen
in Fig.~\ref{pot20Ne.fig}.
Each SUSY-$\beta$ wave function shows the same decrease asymptotically
but its asymptotic normalization coefficient is different from the others
as explained by Eq.~(\ref{asym.eq}). 

\begin{figure}[!ht]
\centering\includegraphics[width=\linewidth]{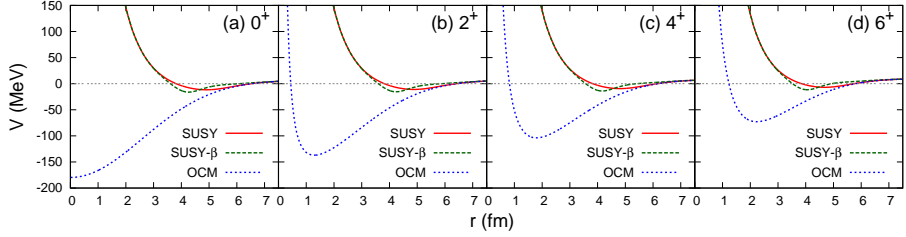}
  \caption{Same as Fig.~\ref{pot20Ne.fig} but
    for the lowest $J^\pi=$ (a) 0$^+$, (b) 2$^+$, (c) 4$^+$, and (d) 6$^+$
    physical states
    of the $\alpha+^{40}$Ca system.}
\label{pot44Ti.fig}
\end{figure}

\begin{figure}[!ht]
  \centering\includegraphics[width=\linewidth]{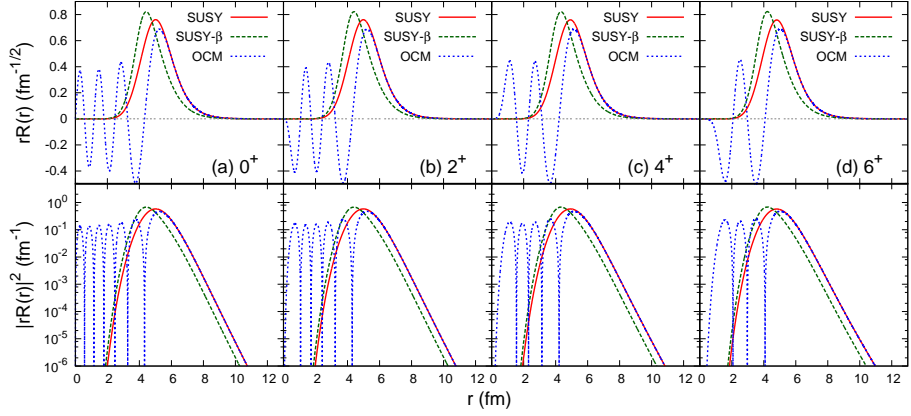}
  \caption{Same as Fig.~\ref{wave20Ne.fig} but 
    with the lowest 
    (a) 0$^+$, (b) 2$^+$, (c) 4$^+$, and (d) $6^+$ physical states
      of $^{44}$Ti.}
  \label{wave44Ti.fig}
\end{figure}

We also display the $\alpha+^{40}$Ca potentials in Fig.~\ref{pot44Ti.fig}.
More singular potentials around the origin are obtained
for the $\alpha+^{40}$Ca system
because the number of forbidden states
($Q=2n_r+l<12$) is larger than that of $^{20}$Ne,
and more transformations
are needed to remove all these forbidden states.
Figure~\ref{wave44Ti.fig} plots the wave functions
and probability distributions
for the lowest $J^\pi=0^+$, $2^+$, $4^+$, and $6^+$ physical
states of $^{44}$Ti. As expected,
two more nodes with respect to the $^{20}$Ne case
appear in the wave function
obtained with the OCM calculation.
Though there is a little difference in the wave functions,
the same discussion given for the $\alpha+^{16}$O system
holds for the asymptotic behavior of the wave function
with the SUSY and SUSY-$\beta$ transformations.

\subsection{Electric-multipole transition densities}
\label{trdens.sec}

\begin{figure}[!ht]
\centering\includegraphics[width=\linewidth]{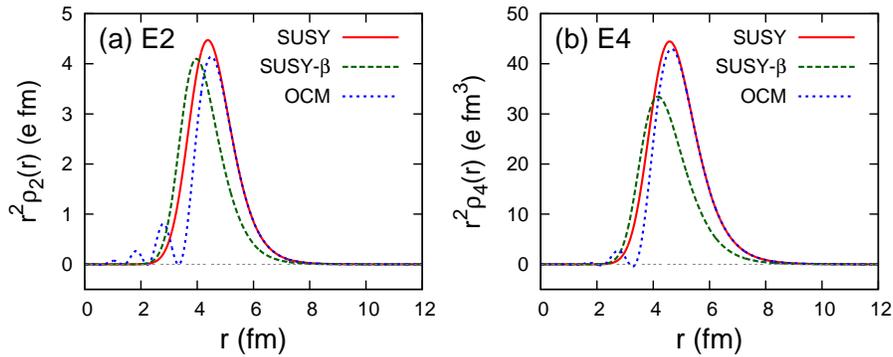}
\caption{Transition densities of the electric- (a) quadrupole ($E2$)
  and (b) hexadecapole ($E4$) operators
  of $^{20}$Ne from the $0^+$ ground state to the yrast
  $2^+$ and $4^+$, respectively,
  obtained with the SUSY transformed and original deep (OCM)
      potentials.}
\label{trdens20Ne.fig}
\end{figure}

In order to clearly see where the electric
multipole transitions occur,
we evaluate the transition densities,
which show the spatial distribution
of the electric-multipole ($E\lambda$) transition matrix elements, defined by
\begin{align}
  \rho_{\lambda}(E,r)=
  \int d\hat{\bm{r}}\,[\Psi_{\lambda\mu}(E,\bm{r})]^*
  \mathcal{M}_{\lambda\mu}\Psi_{00}(E_0,\bm{r})
\end{align}
with the relation to the transition matrix element as
\begin{align}
  \int_0^\infty dr\, r^{2}\rho_{\lambda}(E,r)&=
  \frac{1}{\sqrt{2\lambda+1}}
  \left<\Psi_\lambda(E)\|\mathcal{M}_{\lambda}\|\Psi_0(E_0)\right>.
\end{align}
Figure~\ref{trdens20Ne.fig} plots
the $E\lambda$ transition densities from the ground state
to the positive-parity bound $J^\pi=2^+$ and
  $4^+$ states of $^{20}$Ne.
As expected from
the behavior of the phase-equivalent
wave functions shown in Fig.~\ref{wave20Ne.fig},
the transition densities generated
from the SUSY and original deep potentials
are almost identical beyond $\sim 5$\,fm,
whereas the SUSY-$\beta$ result
is quite different from the others.
Due to the $r^{\lambda}$ factor
in the electric-multipole operators,
the difference below $\sim 5$\,fm
becomes smaller with higher $\lambda$.
Here we can see the reason
why the SUSY-$\beta$ gives the smaller $E4$ values
listed in Table~\ref{phys20Ne.tab}.
We remark that
the SUSY-$\beta$ transformation reduces the
amplitude of the wave function
in the asymptotic regions shown in Fig.\ref{wave20Ne.fig},
and the contributions of the $E4$ transitions
are dominated by the outer regions of the wave function.
Therefore, the $E4$ values become
much smaller then those of the OCM by
the SUSY-$\beta$ transformation.

\subsection{Tests in electric-multipole strength functions
  and cluster sum rules}

To further study the possible modifications of observables
by the SUSY transformations,
we calculate electric-multipole ($E\lambda$)
strength functions.
Here we do not show the results of $^{44}$Ti because
the same discussions as the $^{20}$Ne case can be made.
The $E\lambda$ strength function is defined by
\begin{align}
  S(E\lambda;E)=\mathcal{S}_{f\mu} \left|\left<\Psi_{\lambda\mu}(E_f)|
  \mathcal{M}_{\lambda\mu}|\Psi_{00}(E_0)\right>\right|^2 \delta(E_f-E),
\end{align}
where $\mathcal{S}_{f\mu}$ denotes a sum over the $2\lambda+1$ $\mu$ values and
the final state energy $E_f$.
The final state continuum wave function is normalized as
\begin{align}
\int [\Psi_{\lambda\mu}(E^\prime,\bm{r})]^*\Psi_{\lambda\mu}(E,\bm{r})d\bm{r}
  =\delta(E-E^\prime),
\end{align}
which is practically achieved
by connecting the numerical solution
to the asymptotic Coulomb wave function
with the wave number $k=\sqrt{2m E}/\hbar$ according to
\begin{align}
  \chi(r) \to \sqrt{\frac{2m}{\pi \hbar^2 k}}
  [\cos{\delta_J} F_J(kr) + \sin{\delta_J} G_J(kr)],
  \quad (r\to \infty)
\end{align}
where $\delta_J$ is a phase shift,
$F_J (G_J)$ is a regular (irregular)
Coulomb function~\cite{Abramowitz}.

\begin{figure}[!ht]
\centering\includegraphics[width=\linewidth]{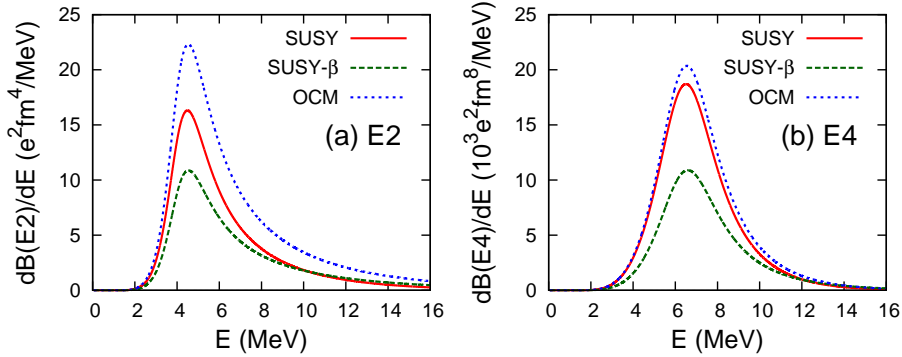}
\caption{Transition strengths of 
  (a) electric-quadrupole ($E2$) and (b) hexadecapole ($E4$)
  operators of $^{20}$Ne
  obtained with the SUSY transformed
    and original deep (OCM) potentials.}
\label{strength20Ne.fig}
\end{figure}

Figure~\ref{strength20Ne.fig} plots
the $E2$ and $E4$ strength functions of $^{20}$Ne
As expected, the OCM and SUSY results become closer
with increasing multipolarity $\lambda$ because
contributions of the asymptotic regions
become more important.
We also calculate the monopole strength function
with the operator $M_{00}=\frac{A_1A_2}{A}r^2$
which allows us to populate the nodal excited $0^+$ states.
Since the operator is proportional to $r^2$,
the behavior of the
transition strengths is similar to that of $E2$
shown in Fig.~\ref{strength20Ne.fig}(a).

\begin{figure}[!ht]
\centering\includegraphics[width=\linewidth]{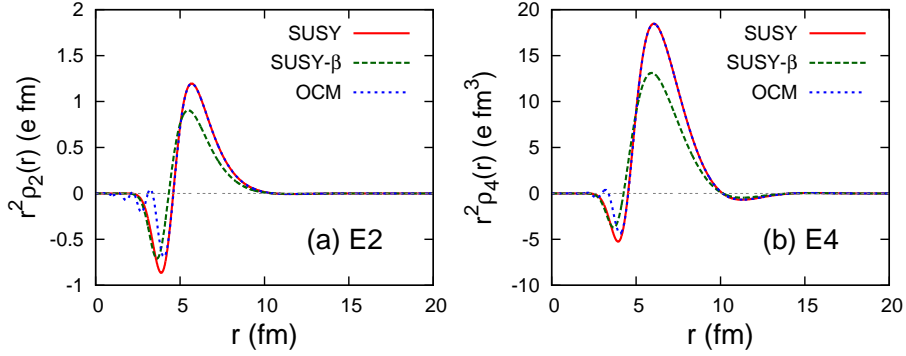}
\caption{Same as Fig.~\ref{trdens20Ne.fig} but
  to the higher nodal states which show the second
  largest transition strengths for each $J^\pi$.}
\label{trdens20Ne2.fig}
\end{figure}

Figure~\ref{trdens20Ne2.fig} plots the transition
densities at the peak positions
for the higher nodal $J^\pi=2^+$ and $4^+$ states
that were observed as broad resonances~\cite{Tilley98}.
The transition densities obtained
with the original deep and SUSY potentials
are identical beyond $\sim 5$\,fm.
The difference in the transition densities is only found
in the internal regions below $\sim 5$\,fm
and becomes smaller with increasing $\lambda$.
In the previous subsection, we discussed the case
where the final state is bound or a sharp resonance.
In this case, the overlap of the initial wave function
multiplied by the $E\lambda$ operator is still large
even at short distances.
Here we discuss transitions to the continuum or
a broad resonant state.
The effects of the internal wave function is much smaller
because the final state wave function is more extended
and displays less amplitude in the internal regions.
The SUSY-$\beta$ transformation always gives smaller strengths than the others
because it reduces the amplitude of the asymptotic wave function.

Finally, we discuss the electric-multipole sum rules.
The non-energy-weighted-cluster sum rules of
the $E\lambda$ transitions are defined by
\begin{align}
  m_0(E\lambda)=\int_{-\infty}^\infty S(E\lambda,E^\prime)\,dE^\prime=
  m_0^{\rm A}(E\lambda)+m_0^{\rm F}(E\lambda)
\end{align}
with the calculable expressions
\begin{align}
m_0(E\lambda)&=
\sum_{\mu}\left<\Psi_{00}(E_0)\right|\mathcal{M}_{\lambda\mu}^\dagger\mathcal{M}_{\lambda\mu}\left|\Psi_{00}(E_0)\right>,\label{SRA.eq}\\
m_0^{\rm F}(E\lambda)&=
\sum_{i=1}^{n}\sum_\mu\left|
\left<\phi^{\rm F}_{\lambda\mu}(E^{\rm F}_i)\right|\mathcal{M}_{\lambda\mu}\left|\Psi_{00}(E_0)\right>\right|^2,
\label{SRB.eq}
\end{align}
where $\phi^{\rm F}_{\lambda\mu}(E_i^{\rm F})$ denotes the wave function
of the forbidden state with the energy $E_i^{\rm F}$.
Note that the sum rule is generalized
and consists of contributions from allowed ($m_0^{\rm A}$)
and forbidden ($m_0^{\rm F}$) states.
It is obvious that $m_0^{\rm F}$ is zero
with use of the SUSY transformed potentials.
Some strengths are distributed
to the redundant bound states when the deep potential is employed.

Table~\ref{sumrule20Ne.tab} lists
the sum-rule values with the $E2$ and $E4$ operators
obtained by the different potential models.
Those values are dominated by $m_0^{\rm A}$.
The $m_0^{\rm F}$ contributions to the total sum rule
are small.
To better quantify the modifications
by the SUSY transformations, we also list the ratios
to the value obtained by the OCM calculation in parentheses.
The $m_0^{\rm A}$ values with the SUSY potential
approach unity with increasing $\lambda$
due to $r^{2\lambda}$ factor in Eq.~(\ref{SRA.eq}).
In contrast, the sum-rule ratios with the SUSY-$\beta$ potential
decrease with increasing multipolarity
because the contributions from the asymptotic regions
become more important with higher multipolarity
and the amplitude of the asymptotic wave function
becomes smaller than that of the original one
as already seen in Fig.~\ref{wave20Ne.fig}.

\begin{table}[!ht]
  \caption{Non-energy-weighted-cluster-electric-multipole ($E\lambda$)
    sum rules of $^{20}$Ne in units of W.u..
    Values in parentheses are the ratios to the value
    obtained by the OCM.}
  \label{sumrule20Ne.tab}
  \centering
  \begin{tabular}{ccccccccc}
\hline
&&&\multicolumn{1}{c}{SUSY}&&\multicolumn{1}{c}{SUSY-$\beta$}&&\multicolumn{1}{c}{OCM}\\
\hline
$E2$&$m_0$        &&{ 133.4} (1.13)&& 98.9 (0.835) && 118.5 \\
    &$m_0^{\rm A}$&& 133.4 (1.18)  && 98.9 (0.873) && 113.2 \\  
    &$m_0^{\rm F}$&&--          &&--            && 5.26\\
$E4$&$m_0$        && 810.4 (1.04) && 471.1 (0.606) && 777.8 \\   
    &$m_0^{\rm A}$&& 810.4 (1.06) && 471.1 (0.617) && 763.3 \\ 
    &$m_0^{\rm F}$&&--          &&--            && 14.6 \\
\hline
\end{tabular}
\end{table}

\section{Conclusion}
\label{conclusion.sec}

Deep potentials can simulate the interaction of two clusters 
  but are practically hard to apply to
  few-body models of multi-cluster systems. 
It is possible to replace these potentials by phase-equivalent shallow 
potentials with the same physical bound states, which are simpler to use. 
The wave functions are however modified. 
In this work, we have quantified this effect with physical properties, 
i.e., electric transition probabilities.

Towards the application of a macroscopic cluster model to 
systems involving heavy clusters,
we have introduced phenomenological deep $\alpha+^{16}$O
and $\alpha+^{40}$Ca potentials
which accommodate several redundant bound states.
We found that,
in order to understand the excitation spectrum as well as
the electric-quadrupole transitions of $^{20}$Ne and $^{44}$Ti,
it is essential for the inter-cluster potential
to reproduce the ground-state energy
and root-mean-square (rms) radius.

Since it is hard to apply this potential directly
to multi-cluster systems, we remove
the unphysical forbidden states by using
supersymmetric (SUSY) transformations.
We have tested the relative wave functions generated
by the SUSY transformed potentials
in comparison with the ones generated by
the original deep potentials.
Though the SUSY transformed potentials give exactly the
same phase shift as the original one, some observables
which are sensitive to the wave function
at short distances (e.g., the rms radius and electric-quadrupole
transition probabilities) are unfavorably modified.
Therefore, we introduce another SUSY transformation (SUSY-$\beta$)
involving one arbitrary parameter which is determined
in such way so as to reproduce the rms radius obtained
by the original wave function.

To quantify the modification caused by the SUSY transformations,
we study the electric-multipole ($E\lambda$) transitions.
The modification due to the standard SUSY transformation
becomes relatively smaller with increasing
multipolarity $\lambda$ due to the $r^\lambda$ factor
in the $E\lambda$ operator.
By applying the SUSY-$\beta$ transformation which keeps the rms radius invariant,
the initial value of the $E2$ transition strengths is recovered.
However, the observables with higher multipolarity, e.g., $E4$,
are too strongly reduced because the single arbitrary parameter included in
the transformation reduces the amplitude
of the asymptotic wave function.

The SUSY prescription offers a phase-equivalent
singular-shallow potential consistent with the deep potential.
Such a potential is advantageous for
descriptions of multi-cluster systems.
However, we should keep in mind that some observables
are possibly modified by the transformation.
The standard SUSY transformation modifies
observables which are sensitive
to the wave functions at short distances.
The modification may be partly recovered by 
another SUSY transformation modifying a bound state,
i.e.\ we can ``design'' the singular-shallow potential.
But again, one needs to care
about the modification of the other observables
that are sensitive to the amplitude of the asymptotic regions.
To further test the potential models presented in this paper,
it is interesting to apply them to
macroscopic three-cluster systems such as
$^{21}$Ne ($^{21}$Na)$=^{16}$O$+\alpha+n$ ($p$),
and $^{24}$Mg$=^{16}$O$+\alpha+\alpha$
systems as well as systems involving
$\alpha$ and $^{40}$Ca clusters.
These applications will verify the validity of
  the potential models through comparison with experimental data.

\section*{Acknowledgment}

W.H. acknowledges an International Promotion Support
from the Department of Physics, Hokkaido University,
that allowed him to visit the 
Universit\'e Libre de Bruxelles (ULB) in February 2017.
He is also grateful to P. Descouvemont for the hospitality
during the stay at ULB.
This work was in part supported
  by JSPS KAKENHI Grant Numbers 18K03635 and
  18H04569.


\section*{References}
  
\begin{thebibliography}{9}
\bibitem{Ikeda68} K. Ikeda, N. Takigawa, H. Horiuchi,
  The Systematic Structure-Change into the Molecule-like Structures
  in the Self-Conjugate $4n$ Nuclei,
  Prog. Theor. Phys. Suppl. {\bf E68} (1968) 464.
\bibitem{Fujiwara80}
  Y. Fujiwara, H. Horiuchi, K. Ikeda, M. Kamimura, K. Kat\=o,
  Y. Suzuki,  E. Uegaki,
  Chapter II. Comprehensive Study of Alpha-Nuclei,
  Prog. Theor. Phys. Suppl. No. {\bf 68} (1980) 29. 
\bibitem{Michel98} F. Michel, S. Ohkubo,  G. Reidemeister,
  Local Potential Approach to the Alpha-Nucleus Interaction and Alpha-Cluster
  Structure in Nuclei,
  Prog. Theor. Phys. Suppl. {\bf 132}, (1998) 7, and references therein.
\bibitem{Cluster}
  Clusters in Nuclei, Volume 1 (2010);
  2 (2012); and 3 (2014), edited by C. Beck,
  Springer (Berlin), and references theirin.
\bibitem{Kamimura81} M. Kamimura,
  Transition densities between the $0_1^+$, $2_1^+$, $4_1^+$, $0_2^+$,
  $2_2^+$, $1_1^-$ and $3_1^-$ states in $^{12}$C derived from the
  three-alpha resonating-group wave functions,
  Nucl. Phys. {\bf A351} (1981) 456.
\bibitem{Suzuki76a} Y. Suzuki,
  Structure Study of $T=0$ States in $^{16}$O by $^{12}$C$+\alpha$
  Cluster-Coupling Model. I,
  Prog. Theor. Phys. {\bf 55} (1976) 1751.
\bibitem{Suzuki76b} Y. Suzuki,
  Structure Study of $T=0$ States in $^{16}$O by $^{12}$C$+\alpha$
  Cluster-Coupling Model. II,
  Prog. Theor. Phys. {\bf 56} (1976) 111.  
\bibitem{Hoyle} F. Hoyle,
  On Nuclear Reactions Occurring in Very Hot STARS.I.
  the Synthesis of Elements from Carbon to Nickel,
  Astrophys. J. Suppl. Ser. {\bf 1} (1954) 121.
\bibitem{RGM1} J.A. Wheeler,
  Molecular Viewpoints in Nuclear Structure,
  Phys. Rev. {\bf 52} (1937) 1083.
\bibitem{RGM1b} J.A. Wheeler,
  On the Mathematical Description of Light Nuclei by the Method of
  Resonating Group Structure,
  Phys. Rev. {\bf 52} (1937) 1107.
\bibitem{RGM2} K. Wildermuth and W. McClure,
  Cluster Representation of Nuclei,
  Springer Tracts in Modern Physics {\bf 41} (1966) 1.
\bibitem{Itagaki95} N. Itagaki, A. Ohnishi,  K. Kat\=o,
  Microscopic $\alpha$-Cluster Model for $^{12}$C and $^{16}$O Based on
  Antisymmetrized Molecular Dynamics: Consistent Understanding of
  the Binding Energies of $^{12}$C and $^{16}$O,
  Prog. Theor. Phys. {\bf 94} (1995) 1019.
\bibitem{OCM1} S. Saito,
Effect of Pauli Principle in Scattering of Two Clusters,
  Prog. Theor. Phys. {\bf 40} (1968) 893.
\bibitem{OCM2} S. Saito,
  Interaction between Clusters and Pauli Principle,
  Prog. Theor. Phys. {\bf 41} (1969) 705.
\bibitem{OCM3} S. Saito,
  Chapter II. Theory of Resonating Group Method and Generator Coordinate
  Method, and Orthogonality Condition Model,
  Prog. Theor. Phys. Suppl. {\bf 62} (1977) 11.  
\bibitem{BFW} B. Buck, H. Friedrich,  C. Wheatley,
Local potential models for the scattering of complex nuclei,
  Nucl. Phys. {\bf A275} (1977) 246.
\bibitem{Kukulin} V.I. Kukulin and V.N. Pomerantsev,
The orthogonal projection method in scattering theory,
  Ann. Phys. {\bf 111} (1978) 330.
\bibitem{Theeten06} M. Theeten, D. Baye,  P. Descouvemont,
Comparison of local, semi-microscopic, and microscopic three-cluster models,
  Phys. Rev. C {\bf 74} (2006) 044304.
\bibitem{Funaki08} Y. Funaki, T. Yamada, H. Horiuchi,
  G. R\"opke, P. Schuck,  A. Tohsaki,
  $\alpha$-Particle Condensation in $^{16}$O Studied with a Full Four-Body
  Orthogonality Condition Model Calculation,
  Phys. Rev. Lett. {\bf 101} (2008) 082502. 
\bibitem{Hiyama10} E. Hiyama, M. Kamimura, Y. Yamamoto,  T. Motoba,
  Five-Body Cluster Structure of the Double-$\Lambda$ Hypernucleus
  $^{11}_{\Lambda\Lambda}$Be,
  Phys. Rev. Lett. {\bf 104} (2010) 212502.
\bibitem{Horiuchi14} W. Horiuchi and Y. Suzuki,
  Correlated-basis description of $\alpha$-cluster and delocalized
  $0^+$ states in $^{16}$O,
  Phys. Rev. C {\bf 89} (2014) 011304(R).
\bibitem{AB} S. Ali and A.R. Bodmer,
  Phenomenological $\alpha$-$\alpha$ potentials,
  Nucl. Phys. {\bf 80} (1966) 99.
\bibitem{Tursunov03} E.M. Tursunov, D. Baye,  P. Descouvemont,
  Comparative variational studies of $0^+$ states in three-$\alpha$ models,
  Nucl. Phys. {\bf A723} (2003) 365. 
\bibitem{Ishikawa13} S. Ishikawa,
Three-body calculations of the triple-$\alpha$ reaction,
  Phys. Rev. C {\bf 87} (2013) 055804. 
\bibitem{Suno15} H. Suno, Y. Suzuki,  P. Descouvemont,
  Triple-$\alpha$ continuum structure and Hoyle resonance of $^{12}$C
  using the hyperspherical slow variable discretization,
  Phys. Rev. C {\bf 91} (2015) 014004.
\bibitem{Baye87}
  D. Baye,
Supersymmetry between deep and shallow nucleus-nucleus potentials,
  Phys. Rev. Lett. {\bf 58} (1987) 2738.
\bibitem{Baye04} D. Baye and J.-M. Sparenberg,
Inverse scattering with supersymmetric quantum mechanics,
  J. Phys. A: Math. Gen. {\bf 37} (2004) 10223.
\bibitem{Baye14} D. Baye, J.-M. Sparenberg, A. M. Pupasov,  B. F. Samsonov,
  Single- and coupled-channel radial inverse scattering
  with supersymmetric transformations,
  J. Phys. A: Math. Theor. {\bf 47} (2014) 243001.
\bibitem{Swan63} P. Swan,
Asymptotic phase-shifts and bound states for two-body central interactions,
  Nucl. Phys. {\bf 46} (1963) 669.
\bibitem{Swan68} P. Swan,
  The exclusion principle and equivalent potentials
  for scattering of complex neutral particles,
  Ann. Phys. NY {\bf 48} (1968) 455.  
\bibitem{Suzuki08} Y. Suzuki, H. Matsumura, M. Orabi, Y. Fujiwara,
  P. Descouvemont, M. Theeten,  D. Baye,
  Local versus nonlocal $\alpha\alpha$ interactions in a $3\alpha$ description
  of $^{12}$C,
  Phys. Lett. {\bf B659} (2008) 160. 
\bibitem{Pinilla11} E.C. Pinilla, D. Baye, P. Descouvemont,
  W. Horiuchi,  Y. Suzuki,
Tests of the discretized-continuum method in three-body dipole strengths,
  Nucl. Phys. {\bf A865} (2011) 43.
\bibitem{Baye87b} D. Baye,
Phase-equivalent potentials from supersymmetry,
  J. Phys. A: Math. Gen. {\bf 20} (1987) 5529. 
\bibitem{Sukumar85} C.V. Sukumar,
Supersymmetric quantum mechanics of one-dimensional systems,
  J. Phys. A: Math. Gen. {\bf 18} (1985) 2917.  
\bibitem{Horiuchi14b} W. Horiuchi and Y. Suzuki,
Tests of a Deformable Core Plus Few-Nucleon Model,
  Few-Body Syst. {\bf 55} (2014) 121.
\bibitem{Baye18} D. Baye and E.M. Tursunov,
  Isospin-forbidden electric-dipole capture and the $\alpha(d,\ \gamma) ^6$Li
  reaction,
  arXiv: 1710.06352.
\bibitem{Kruppa90} A.T. Kruppa and K. Kat\=o,
  Resonances in Complex-Scaled Orthogonality Condition Model
  of Nuclear Cluster System,
  Prog. Theor. Phys. {\bf 84} (1990) 1145.  
\bibitem{Abramowitz}
  M. Abramowitz and I.A. Stegun,
  Handbook of Mathematical Functions with
  Formulas, Graphs,  Mathematical Tables, Dover, Mineola, NY (1970).
\bibitem{Angeli13}
  I. Angeli and K.P. Marinova,
  Table of experimental nuclear ground state charge radii: An update,
  Atomic Data and Nuclear Data Tables {\bf 99} (2013) 69.
\bibitem{Wang17a} W.J. Huang, G. Audi, M. Wang, F.G. Kondev,
  S. Naimi,  X. Xu,
  The AME2016 atomic mass evaluation (I). Evaluation of input data;
  and adjustment procedures,
  Chin. Phys. C {\bf 41} (2017) 030002.
\bibitem{Wang17b}  M. Wang, G. Audi, F.G. Kondev,
  W.J. Huang, S. Naimi,  X. Xu,
  The AME2016 atomic mass evaluation (II). Tables, graphs and references,
  Chin. Phys. C {\bf 41} (2017) 030003.
\bibitem{Horiuchi68} H. Horiuchi and K. Ikeda,
  A Molecule-like Structure in Atomic Nuclei of $^{16}$O$^*$ and $^{10}$Ne,
  Prog. Theor. Phys. {\bf 40} (1968) 277. 
\bibitem{Tilley98} D.R. Tilley, C.M. Cheves,
  J.H. Kelley, S. Raman,  H.R. Weller,
  Energy levels of light nuclei, $A=20$,
  Nucl. Phys. {\bf A636} (1998) 249.
\bibitem{BM} A. Bohr and B.R. Mottelson,
  Nuclear Structure, Vol. I, W.A. Benjamin, New York (1975).
\bibitem{Chen11} J. Chen, B. Singh, J.A. Cameron,
  Nuclear Data Sheets for $A=44$,
  Nucl. Data Sheets {\bf 112} (2011) 2357.
\bibitem{Kimura06} M. Kimura and H. Horiuchi,
  Coexistence of cluster structure and superdeformation in $^{44}$Ti,
  Nucl. Phys. {\bf A767} (2006) 58.
\bibitem{Whittaker}
  E.T. Whittaker, G.N. Watson, A Course of Modern Analysis,
  Fourth edition, Cambridge Univ. Press, Cambridge (1927).
\end{thebibliography}
\end{document}